\documentclass[aps,pre,reprint,groupedaddress,amsmath,amssymb,byrevtex,floatfix]{revtex4-1}

\usepackage{graphicx}
\usepackage{color}
\definecolor{colortodo}{RGB}{0,0,0}
\newcommand{\red}[1]{{\color{colortodo}#1}}

\begin{document}

\title{Tuning the distance to equipartition by controlling the collision rate in a driven granular gas experiment}

\author{Gustavo Castillo} \affiliation{Instituto de Ciencias de la Ingiener\'ia, Universidad O'Higgins, 2841959 Rancagua, Chile}
\author{Simon Merminod} \affiliation{Martin A. Fisher School of Physics, Brandeis University, Waltham, MA 02453, USA }
\author{Eric Falcon} \affiliation{MSC, Universit\'e de Paris, Universit\'e Paris Diderot, CNRS (UMR 7057), 75013 Paris, France}
\author{Michael Berhanu} \affiliation{MSC, Universit\'e de Paris, Universit\'e Paris Diderot, CNRS (UMR 7057), 75013 Paris, France}
\email{michael.berhanu@univ-paris-diderot.fr}

\begin{abstract} 
In a granular gas experiment of magnetized particles confined in a thin layer, the rate of dissipative collisions is tuned by adjusting the amplitude of an external magnetic field. The velocity statistics are analyzed using the dynamic and static structure factors of transverse velocity modes. Using the fluctuating hydrodynamics theory we measure the deviation from kinetic energy equipartition in this out-of-equilibrium system as a function of the dissipative collision rate. When the collision rate is decreased, the distance to equipartition becomes smaller meaning that the dynamical properties of this granular gas approach by analogy those of a molecular gas in thermal equilibrium.
\end{abstract}

\maketitle

\section{Introduction}
Statistical mechanics succeeds in predicting the macroscopic states of systems composed of many interacting particles, mainly when these systems are in thermal equilibrium. Such systems have a reversible dynamics and do not dissipate energy. In contrast, for out-of-equilibrium systems like turbulent flows, biological living systems, active fluids or electrical circuits, only few general results statistically describe the nonequilibrium steady states~\cite{Derrida2007}, in which energy must be continuously injected to compensate for energy dissipation~\cite{Bertin2017}. Among them, granular gases refer to an assembly of athermal macroscopic particles mechanically agitated which undergo dissipative collisions. They are relevant model systems to investigate non-equilibrium steady states theoretically~\cite{Barrat2005,VanNoije1998,Goldhirsch1993}, numerically~\cite{McNamara1998,Moon2004,Aumaitre2006,Opsomer2011} and experimentally~\cite{Olafsen1998,Olafsen1998,Losert1999,Reis2007a,Mujica2016,Noirhomme2018}. Taking into account the inelasticity of collisions and  assuming that the forcing acts as a stochastic noise~\cite{Kadanoff1999,Goldhirsch1999}, kinetic theories propose a method for predicting the large scale behavior of many particle systems. For instance, mode coupling theory models velocity structure factors and predicts long range spatial correlations caused by dissipative collisions~\cite{vanNoijeErnst1999}. More recently, for a driven granular system fluidized by a stochastic bath with friction, the fluctuating hydrodynamics theory~\cite{Gradenigotheo2011} also derives the static velocity structure factors and finds correlation lengths related to energy dissipation, in order to model vibrated granular experiments.  Due to the dissipative collisions, the kinetic energy per particle at large scales called the ``bath temperature" is higher than the one at the particle scale, the ``grain temperature". The energy equipartition is thus violated through the space scales. This approach has been successfully validated in a quasi-two-dimensional experiment of homogeneously driven granular particles~\cite{Gradenigo2011,Puglisi2012}. In that work, the level of dissipation is varied by changing the number of particles per area unit, the area fraction. However, their structure factors remain dominated by collision effects because the particle area fraction must be kept high enough to maintain the validity of the hydrodynamics approach. 

\begin{figure}[t!]
\includegraphics[width=0.95\columnwidth]{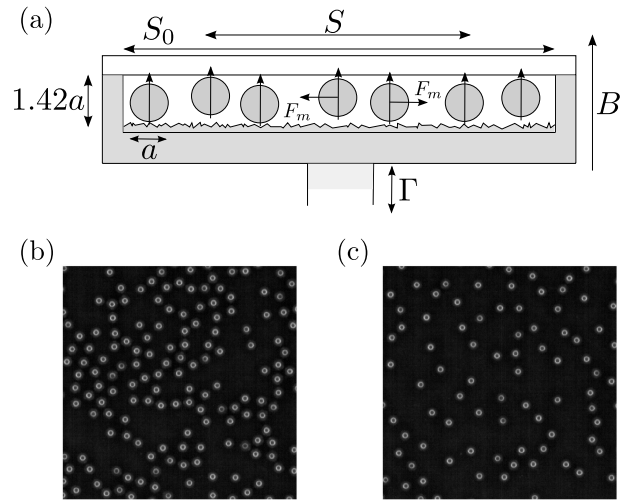}
\caption{(a) Schematic of the experimental setup (not to scale). Magnetized granular spheres of diameter $a=1$~mm are imaged  from the top with a high-speed camera in the region of interest $S$. The spheres are immersed in a transverse magnetic field $B$ and vertically shaken with acceleration $\Gamma$. (b) Snapshot of the experiment for $B\approx 0$~G ($\varepsilon \approx 0$), and (c) for $B=62\,$G ($\varepsilon= 8.80$). Shaker acceleration is $\Gamma=1.6g$ and snapshot size is 17.5 $\times$ 17.5 mm$^2$.}
\label{fig1}
\end{figure}

In a previous work~\cite{Merminod2014}, we introduced a different experimental setup realizing a homogeneously driven quasi two-dimensional granular gas, in which tunable inter-particle repulsive forces have been added by means of an external magnetic field. When these forces are strengthened, the rate of dissipative collisions decreases because collisions are progressively replaced by elastic dipolar interactions. In a range of moderated applied magnetic field, the statistical properties of the granular gas approach those of a molecular gas in thermal equilibrium. To our knowledge, the effect of repulsive interactions on the dissipation rate has been firstly investigated theoretically and numerically by Scheffler and Wolf~\cite{Scheffler2002} for a granular gas of electrically charged particles in ballistic motion. Here, we quantify the distance to kinetic energy equipartition in our experiment throughout the transition from a dissipative granular gas to a quasi-elastic system of particles. Specifically, we use the methods from the fluctuating hydrodynamics theory~\cite{Gradenigotheo2011} to compute the difference between the bath and the grain temperatures to define a distance to equipartition. We observe that once the collision rate vanishes, these two temperatures are nearly equal. Indeed, we show that the deviation from equipartition is caused by energy depletion at small scales due to dissipative collisions.

\begin{figure*}[t!]
{\includegraphics[width=2.1\columnwidth]{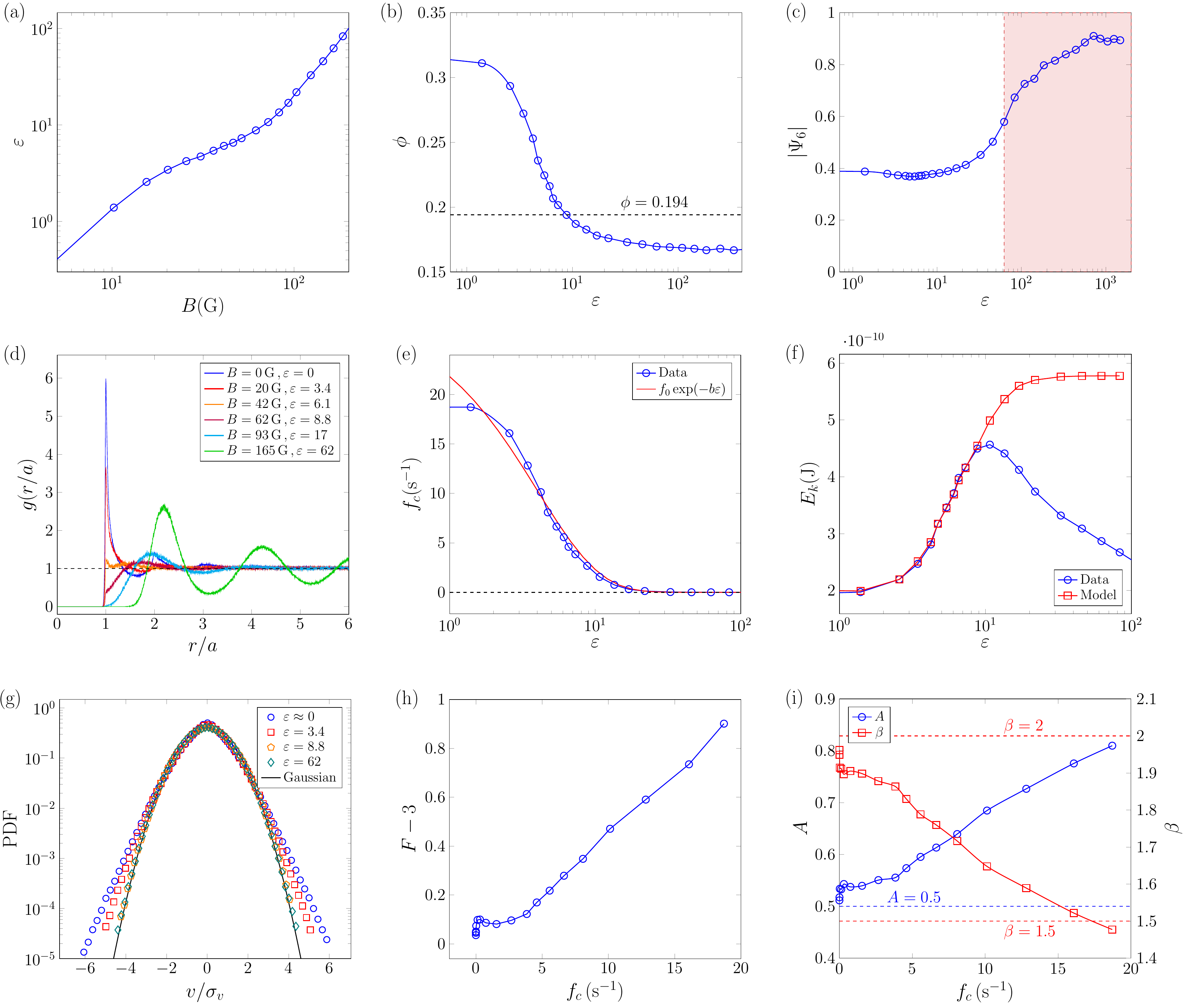}}
\caption{(a) Ratio between magnetic and kinetic energy per particle $\varepsilon =E_m/E_k$ as a function of the applied magnetic field $B$. (b) Area fraction $\phi$ as a function of $\varepsilon$. (c) Hexagonal order parameter $|\Psi_6|$ as a function of $\varepsilon$. Shaded area corresponds to the hexagonal phase (described in~\cite{Merminod2014}). (d) Radial distribution function $g(r)$ for various values of $\varepsilon$. (e) Collision rate or collision frequency between particles $f_c$ as a function of $\varepsilon$. Red line shows the modeled exponential decay of the collision rate: $f_c = f_0\,\exp(-b\,\varepsilon)$, with $f_0 = 28\,\mathrm{s}^{-1}$ and $b=0.25$. (f) Kinetic energy experimentally measured ($E_k$, blue) and modeled ($E_k^{th}$ from Eq.~\ref{Ekmod}, red) as a function of $\varepsilon$. The model relates the increase of $E_k$ to the decrease of $f_c$. (g) Probability distribution function (PDF) of the particle horizontal velocities $v$, normalized by the standard deviation $\sigma_v$ for selected values of $\varepsilon$. The black line corresponds to the Gaussian distribution. Both spatial coordinates ($x$ and $y$) are statistically equivalent and are used to compute the velocity PDF. (h) Kurtosis or flatness of the velocity distributions ($F=\langle v^4 \rangle /\sigma_v^4 $) minus $3$ as a function of the collision rate $f_c$ for $1 < \varepsilon<100$. (i) Fit parameters of modeled velocity PDF, $A$ and $\beta$ (see Eq.~\ref{strechexpo}) versus the collision rate $f_c$. Blue dashed line, value $A=1/2=0.5$ expected for a Gaussian distribution. Red dashed lines, $\beta=2$ expected for a Gaussian distribution and $\beta=3/2=1.5$ for a homogeneously driven granular gas.}
\label{fig2}
\end{figure*}

\section{Transition towards a collisionless granular gas}

First, we recall the features of the experimental device~\cite{Merminod2014,Merminod2015,MerminodPhd} whose schematic is given in Fig.~\ref{fig1} (a). An assembly of $2000$ soft magnetic spheres of diameter $a=1$\,mm and of mass $m=4.07\times10^{-3}$\,g is confined in a square cell of surface $S_0=90\times 90$\,mm$^2$ with a vertical gap of $1.42\,a$. The cell is vertically vibrated at a frequency $f_v=300\,$Hz with a r.m.s. acceleration $\Gamma=(2\pi f_v)^2\,A_0/g=1.6 \,g$,where $A_0$ is the amplitude of vibration and $g$ is the gravitational acceleration. \red{Particles perform Brownian-like motion in the horizontal plane due to the roughness of the bottom surface of the cell, whose r.m.s. rugosity is measured to be $20\,\mu$m}. By imaging with a high-speed camera (Phantom V10) the area $S$ (\red{$50.36 \times 50.36$\,mm$^2$}) through a transparent \red{and smooth} lid, the trajectory of each particle is reconstructed in two-dimensions, in the horizontal plane. \red{The ratio of the surface of the imaged area $S$ to the surface of the entire cell $S_0$ reads $S/S_0 =0.315$.} When immersed in the external vertical magnetic field $B$, each sphere behaves as an induced dipole. In a first approximation, two particles whose centers are separated by a distance $r_{i,j}$ interact according to the repulsive potential $U_{i,j}=\frac{4\pi}{\mu_0}\,B^2 \frac{(a/2)^6 }{r_ {i,j}^3}$ \cite{Jackson1999,MerminodPhd}, where $\mu_0$ is the permeability constant. The relevant parameters of our experiments are the area fraction $\phi=\frac{N\pi a^2}{4S}$, with $N$ the average number of spheres imaged in $S$ (here $\phi \approx 0.2$), the mean kinetic energy per particle $E_k=\langle \frac{m}{2N}\sum_{i=1}^N v_i^2 \rangle$ (with $v_i$ the velocity of particle $i$ \red{in the horizontal plane} and the brackets denote time averaging) and the mean magnetic energy per particle $E_m=\langle \,\frac{1}{N}\sum_{i=1}^N \sum_{j=i+1}^N U_{i,j} \rangle $. The dimensionless number $\varepsilon=E_m/E_k$ quantifies the competition between the interaction strength and kinetic energy. $B$ is varied in $[-0.11,430]$\,G, corresponding to $\varepsilon \in [6.5 \times 10^{-4} \,,1.5\times10^3]$.  Each measurement is averaged over five independent realizations. After an equilibration time of $100$\,s, images are acquired using the high-speed camera during $3.85\,$s at a frame rate of $780\,$Hz. Using tracking algorithms~\cite{Shattuck}, the position, trajectory and velocity of each particle \red{in the horizontal plane} are computed in the window of observation ${S}$. For this vertical confinement distance of approximately $1.42\,a$ and area fraction $\phi \approx 0.2$,  it has been shown~\cite{Schockmel2013,Merminod2014,Schockmel2017}, that the dipolar interaction remains purely repulsive and the system can be described as two-dimensional. For larger gap, three-dimensional effects must be taken into account for the interaction potential and the spatial distribution of spheres, leading to a large variety of phases~\cite{Merminod2015,MerminodPhd,Opsomer2019}.
When the external magnetic field is increased, the strength of magnetic interactions quickly overcomes kinetic agitation as shown in the plot of $\varepsilon=E_m/E_k$ as a function of $B$ (Fig.~\ref{fig2} (a)). As $\varepsilon$ is increased, the competition between repulsive interactions and kinetic agitation results in a transition from a granular gas towards a hexagonal crystal. Snapshots of a window inside $S$ are shown without applied magnetic field in Fig.~\ref{fig1} (b) ($\varepsilon \approx 0$) and with a moderate value of $B=62\,$G ($\varepsilon \approx 8.80$) in Fig.~\ref{fig1} (c). In both cases, the assembly of spheres is in a granular gas state, but in the second snapshot the particles do not come into contact anymore. We note also a smaller number of particles in the second case. In Fig.~\ref{fig2} (b), we show indeed a decrease of $\phi$ with $\varepsilon$ in the observation window $S$, which is due to increasing particle repulsion while boundaries are non-repulsive. The crystallization towards a hexagonal crystal is monitored by the sixfold bond-orientational order parameter per particle
\begin{equation}
\Psi^j_6=\frac{1}{n_j}\sum_{k=1}^{n_j}\mathrm{e}^{6 i \theta_{jk}},
\end{equation}
 where $n_j$ is the number of nearest neighbors of particle $j$, and $\theta_{jk}$ is the angle between the neighbor $k$ of particle $j$ and a reference axis. The corresponding global average,
\begin{equation}
|\Psi_6| = \left |\left \langle \frac{1}{N}\sum_{j=1}^N \Psi^j_6\right\rangle\right|,
\end{equation}
where the vertical bars denote a modulus, measures the degree of hexagonal order of the particle assembly. We will refer to $|\Psi_6|$ as the hexagonal order parameter. In Fig.~\ref{fig2} (c) $|\Psi_6|$ is plotted as a function of $\varepsilon$. $|\Psi_6|$ is of order $0.4$ in the granular gas phase, to reach $0.9$ in the hexagonal crystal phase. The transition towards the hexagonal phase is located at $\varepsilon_c \approx 62$, corresponding to a maximal susceptibility, \textit{i.e} the maximal variation 
of the hexagonal order parameter $|\Psi_6|$ to changes of $\varepsilon$. Here, our study is focused on the granular gas phase, thus for $\varepsilon < 62$, \textit{i.e.} $B \lesssim 165$\,G. Although the system remains in a fluid-like phase for this range of $B$, it undergoes important structural changes. In Fig.~\ref{fig2} (d), the radial pair distribution function $g(r)$ (or radial pair correlation function)~\cite{HansenMcdonaldBook,Barrat2003a} is plotted as a function of the center-to-center distance between spheres $r$ for selected values of $\varepsilon$. For $\varepsilon=0$, $g(r)$ displays a strong peak at contact between spheres, \textit{i.e.} $r=a$ due to hard-sphere repulsion. For larger values of $\varepsilon$, the amplitude of this peak decreases due to strengthened inter-particle repulsion. In particular, $g(r)$ becomes nearly flat for $\varepsilon=6.1$ denoting the quasi absence of spatial correlations. When repulsion is further increased, the contacts become unlikely and a depleted zone appears for $r$ slightly larger than $a$. At $\varepsilon=62$, $g(r)$ displays spatial oscillations characteristic of an emerging hexagonal order~\cite{Merminod2014}. For roughly $\varepsilon>5$, the rate of collisions between particles, $f_c$, defined as the average number of distinct events per second for which $r<1.03\,a$, is strongly reduced by the increasingly repulsive interactions (Fig.~\ref{fig2} (e)). It nearly vanishes for $\varepsilon \approx 20$. For even larger $\varepsilon$, the dissipative collisions disappear. Then, particles interact with each other only through magnetic dipolar interactions, which are elastic (\textit{i.e.} conservative). The decrease of $f_c$ with $\varepsilon$ is roughly approximated by a decaying exponential function, $f_0\, \mathrm{exp}(- b\, \varepsilon) $, with $f_0=28\,$s$^{-1}$ and $b=0.25$. To collide two particles must overcome an energy barrier due to magnetic repulsion. Hence, the collision rate should be proportional to an Arrhenius factor $\mathrm{exp}(-\varepsilon)$~\cite{Scheffler2002}. Factor $b$ being different from unity in our experiments may be explained by the exclusion of vertical motion from our analysis or by collective effects if the granular gas is not dilute enough~\cite{Scheffler2002}. Note also that, the variations of $\phi$ and $E_k$ are too limited in this set of measurements to test their expected influence on the collision rate. However, a possible dimensional scaling is $f_c \sim \sqrt{E_k/m}\,\phi\,a^{-1}\,g_ {eq}(r=a)\,\mathrm{exp}(- b\, \varepsilon) $, by assuming that $f_c$ is given by the product of the mean quadratic velocity by the mean free path, whose expression is given for hard-spheres by $ (a\,\sqrt{\pi})/(2\,g_ {eq}(r=a)\,\phi)$~\cite{Puglisi2012} with $g_ {eq}(r=a)$ the value of the equilibrium pair correlation function at contact. 

The measured kinetic energy per particle $E_k$ has a non-monotonous behavior, as a function of $\varepsilon$ (Fig.~\ref{fig2} (f)). For $0< \varepsilon <10$, $E_k$  increases due to the decrease of the collision rate. Then for  $\varepsilon>10$, \textit{i.e.} for stronger magnetic repulsion, $E_k$ significantly decreases. For a granular gas in stationary regime, by balancing energy injection with dissipation~\cite{Aumaitre2006}, the theoretical mean kinetic energy per particle can be written as: 
\begin{equation}
E_k^{th} =  \dfrac{\langle P \rangle}{\left[ (1-r^2)\,f_c + \delta \right]}
\label{Ekmod}
\end{equation}
where $\langle P \rangle$ is the average injected power per particle, $\delta \, E_k^{th}$ is the average dissipation due to the collisions of particles with the bottom and top walls, whereas  $(1-r^2)\,f_c \, E_k^{th}$ is the dissipation caused by the inelastic collisions between particles, and $r=0.9$ is a realistic restitution coefficient~\cite{Opsomer2019}. Using the experimentally measured collision rate $f_c$, the growth of $E_k$ as a function of $\varepsilon$ is well described by Eq.~\ref{Ekmod} with fitted parameters $\langle P \rangle =1.1\times10^{-9}$ W and $\delta=1.9$ (red curve in Fig.~\ref{fig2} (f)). For $\varepsilon \geq 10$, the magnetic repulsion constrains the horizontal motions perpendicular to the applied magnetic field, to favor the vertical motions and thus decreases the effective injected power. Finally, as reported for other quasi-two-dimensional vibrated granular gas experiments~\cite{Olafsen1999,Losert1999,Rouyer2000,Reis2007,Puglisi2012,Merminod2014,Scholz2017}, the distribution of  particle velocities $v$ for $\varepsilon =0$ deviates from the Gaussian distribution expected for a molecular gas in equilibrium (Fig.~\ref{fig2} (g)). When $\varepsilon$ is increased, the distance to the Gaussian distribution decreases due to the diminution of the particle collision rate. Indeed, when the collision rate decreases, the kurtosis of the velocity distribution, or the ``flatness", $F=\langle v^4 \rangle /\sigma_v^4 $, where $\sigma_v$ is the standard deviation of the velocity distribution, approaches the value~$3$ expected for a Gaussian distribution (Fig.~\ref{fig2}~(h)). We note that the velocity distributions are satisfactorily fitted by a stretched exponential function
\begin{equation}
f(v) \propto \mathrm{exp}(-A \vert v/\sigma_v\vert^\beta)\,\, ,
\label{strechexpo}
\end{equation}
with $A$ and $\beta$ varying respectively from $0.81$ to $0.51$ and $1.40$  to $1.97$ as $f_c$ decreases (Fig.~\ref{fig2} (i)). For a homogeneously driven granular gas with dissipative collisions kinetic theory predicts an exponent $\beta=3/2$ for the high energy tail of the velocity PDF~\cite{vanNoijeErnst1999}. In our experiment, \red{the fitted value of $\beta$ is $1.48$ for $f_c=18.7$\,s$^{-1}$ the largest collision rate. $\beta$ is thus close to this theoretical prediction.} In contrast, when $f_c \approx 0$, \red{we find $A=0.520$ and  $\beta=1.94$. These values} approach those expected for a Gaussian distribution $A=1/2$ and $\beta=2$. Therefore, as $B$ is increased and $f_c$ diminishes, the continuous increase of $\beta$ from $1.40$  to $1.97$, shows unambiguously that the shape of the velocity distribution depends on the collision rate in a granular gas. Similarly, the velocity distribution can be fitted using the one-dimensional Sonine poynomial corrections to a Gaussian distribution~\cite{Coppex2003,Brilliantov2006,Reis2007}. The experimental velocity distributions are approximately reproduced using the development at the second order (not shown), with Sonine coefficients $a_1=0$, $a_2$ in the range $[0.01,0.26]$~
\footnote{The velocity distribution in terms of the Sonine polynomial expansion reads as a function of the rescaled velocity $c=v/\sqrt{2\,\sigma_v}$ $f(v)=\pi^{-1/2}\,\mathrm{exp}(-c^2)\,(1+a_1\,S_1(c^2)+a_2\,S_2(c^2))$ at second order, with $S_1(x)=-x+\frac{1}{2}$, $S_2(x)=\frac{1}{2}\,x^2-\frac{3}{2}\,x+\frac{3}{8}$}.
 The coefficient $a_2$ decreases indeed with $f_c$ and the kurtosis verifies accurately $F=3\, (1+a_2)$.

\begin{figure*}[t!]                    
  \begin{center}
     \includegraphics[width=2.1\columnwidth]{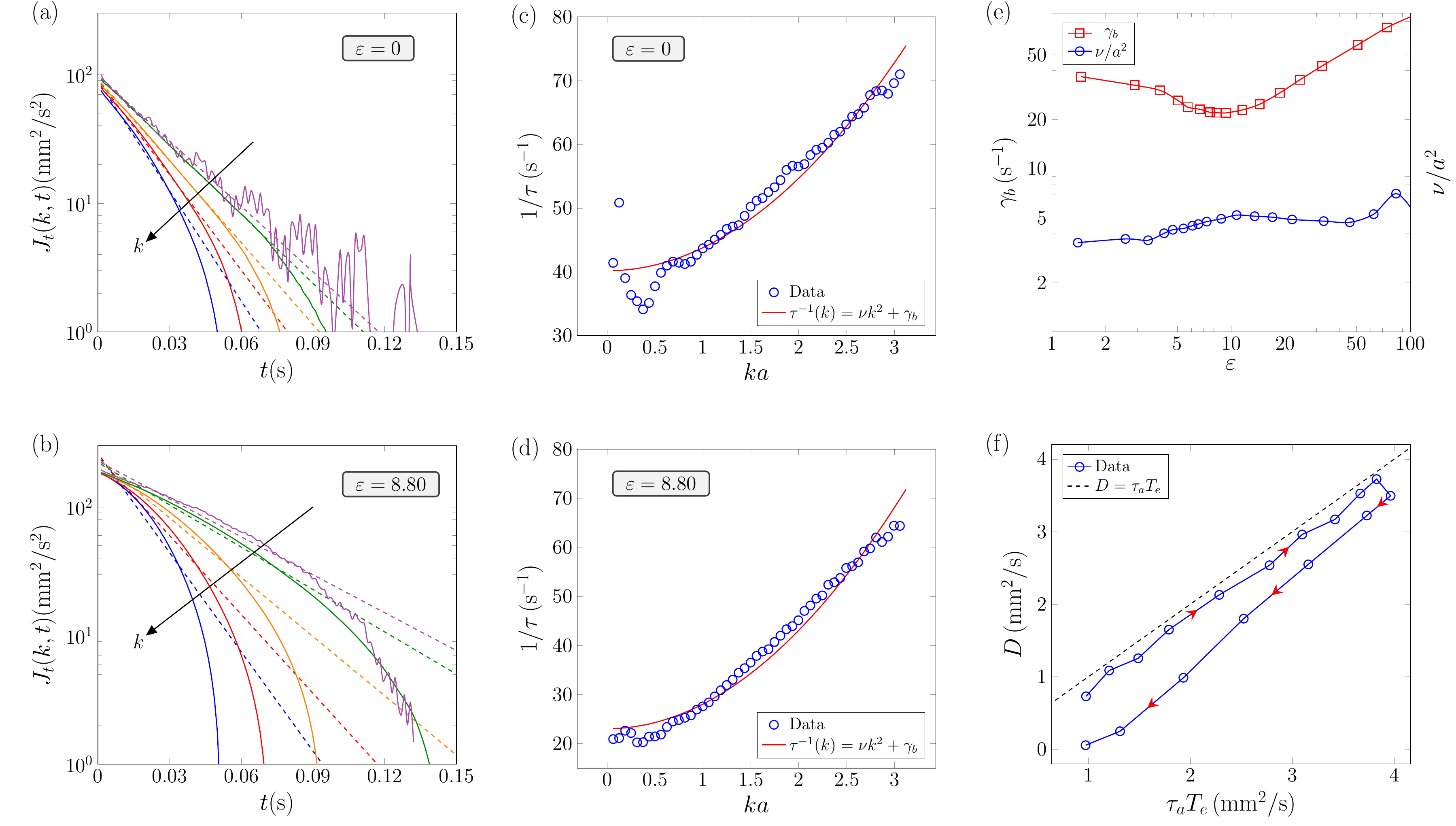}  
     \caption{Temporal decay of the transverse current correlation functions $J_{t}(k,t)$, for selected $k$ increasing from top to bottom for $\varepsilon =0$ (a) and $\varepsilon=8.80$ (b). Dashed lines, exponential fits measuring the dissipative time $\tau(k)$. The behavior is similar for larger values of $\varepsilon$ in the granular gas phase ($\varepsilon \lesssim 50$). (c) Inverse of the fitted typical dissipative time $\tau(k)$ as a function of $k\,a$ for $\varepsilon =0$, and (d) for $\varepsilon=8.80$. (e) Evolution of the friction coefficient $\gamma_b$ and of the rescaled kinematic viscosity $\nu/a^2$ with $\varepsilon$. (f) Self-diffusion coefficient $D$ as a function of $\tau_a \,T_e$, with $\tau_a=\tau (k=1/a)$ and $T_e=E_k/m$. Each data point corresponds to a value of $\varepsilon$ (for $\varepsilon < 62$). Red arrowheads point towards greater $\varepsilon$. Dashed line corresponds to $D = \tau_aT_e$.}
    \label{fig3}            
  \end{center}
\end{figure*}

\section{Effective dissipation parameters}

We showed in previous Section that the dynamical behavior of this assembly of magnetized spheres is strongly controlled by the particle collision rate for moderate values of $\varepsilon$, by studying the statistics of individual velocities. We now investigate the collective dynamics to obtain a different characterization of the dissipative processes at work. For that, velocity correlations can be computed in the spatial Fourier space and analyzed in the framework of linearized hydrodynamics~\cite{HansenMcdonaldBook} which provides effective transport coefficients. A fruitful approach is to compute the dynamical transverse velocity structure factor $J_t (\mathbf{k},t)$~\cite{HansenMcdonaldBook,Gradenigo2011,Puglisi2012,CastilloPhd} (also called the transverse current correlation function): 
$$ J_t (\mathbf{k},t)=\left\langle \frac{1}{N}  \sum_{i,j=1}^N (\mathbf{\hat{k}} \times \mathbf{v}_i) (t) \, (\mathbf{\hat{k}} \times \mathbf{v}_j) (0)\, \mathrm{e}^{i\,\mathbf{k}\, (\mathbf{r}_i (t)-\mathbf{r}_j(0)) }    \right\rangle  $$
where $t$ is the time, $\mathbf{k}$ is the wave vector, $\mathbf{\hat{k}}$ is the unitary vector directed along $\mathbf{k}$, $\mathbf{v}_i$ (resp., $\mathbf{v}_j$) is the velocity vector of particle $i$ (particle $j$), and $\mathbf{r}_i$ (resp., $\mathbf{r}_j$) is the position vector of particle $i$ (particle $j$).
$\langle \, \rangle$ denotes a time average. These structure factors are computed for a stationary forcing \red{for each test value of $\mathbf{k}=[k_x,k_y]$ chosen in a horizontal grid of size $50 \times 50$ in the domain $0.0624 < k_{x,y} <3.12\,$mm$^{-1}$. Each pair $(k_x,k_y)$ is discretized according to $(n_x\pi/L_x,n_y\pi/L_y)$, where $n_x, n_y\in\mathbb{N}$ and $L_x=L_y=50.36$\,mm. A loop and a nested loop over the image numbers perform sweeps of initial times $t=0$ and time lags $t$ of the correlation function. Then, the quantity $ (\mathbf{\hat{k}} \times \mathbf{v}_i) (t) \, (\mathbf{\hat{k}} \times \mathbf{v}_j) (0)\, \mathrm{e}^{i\,\mathbf{k}\, (\mathbf{r}_i (t)-\mathbf{r}_j(0)) }$ is computed by separating the real and imaginary parts for each pair of particles $i$ at time $t$ and $j$ at time $0$. $J_t (\mathbf{k},t)$ is obtained by first ensemble averaging and then time averaging. Assuming isotropy of particle motions, the quadrants for negative $k_x$ or $k_y$ are reconstructed. The angular average is computed as $J_t (k,t)=(2\pi)^{-1} \,\int_0^{2\pi} J_t (\mathbf{k},t)\, \mathrm{d} \theta$, where $k = ||\mathbf{k}||$. $J_t (k,t)$ is finally averaged over five independent runs with identical experimental parameters.}

The typical decay time of the transverse current $J_t(k,t)$ characterizes the dissipative processes at play. In the granular gas phase ($\varepsilon < 62$), the short-time decrease of $J_t(k,t)$ is well approximated by a decaying exponential $\sim \mathrm{e}^{-t/\tau (k)}$, where $\tau (k)$ is the typical life time of an excitation at the scale $k$ (Fig.~\ref{fig3} (a) and (b)). For vibrated granular layers, energy dissipation is often modeled by the combination of a viscous drag and of a Coulomb friction leading to the equation $\tau^{-1}(k)=\nu\,k^2+\gamma_b$~\cite{Puglisi2012,CastilloPhd}, with $\nu$ a kinematic viscosity and $\gamma_b$ a friction coefficient. For each value of $\varepsilon$, parameters $\nu$ and $\gamma_b$ are obtained by fitting this equation to the measured values of $\tau^{-1}(k)$ (Fig.~\ref{fig3}~(c) and~(d)). Note that such a modeling of energy dissipation becomes invalid for high $\varepsilon$, when the hexagonal phase is reached. \red{Moreover, the fit quality worsens at small $k\,a$, where the statistical convergence is lesser and finite size effects may interfer.} Parameters $\nu /a^2$ and $\gamma_b$ are plotted as a function of $\varepsilon$ in the granular gas phase ($0<\varepsilon<62)$ in Fig.~\ref{fig3} (e). For this set of experiments, dissipation is dominated by friction. A characteristic length $\xi = \nu/\gamma_b$ of order $0.3\,$mm can be defined. Surprisingly, this value is significantly smaller than the one found in Puglisi et al.~\cite{Puglisi2012}. Important differences between their system and ours include that their particles do not remotely interact and are more strongly agitated, and that their experimental cell has no lid.The fluctuating hydrodynamics theory interprets this length as a spatial correlation length of excitations~\cite{Gradenigotheo2011,Gradenigo2011,Puglisi2012}. 

The relation between particle diffusion and dissipation is studied as follows. The self-diffusion coefficient, $D$, is obtained by fitting the mean-squared displacements of particles using the equation $\langle [\mathbf{r}_i (t) - \mathbf{r}_i (t=0) ]^2\rangle = 4Dt$. Standard diffusion remains valid in the granular gas phase until approximately $\varepsilon < 50$. By analogy with the Einstein relation~\footnote{According to the Einstein relation, the self-diffusion coefficient $D$ of a molecular fluid writes $ D=k_B\,T/\Lambda $, with here $k_B$ the Boltzmann constant, $T$ the thermodynamic temperature and $\Lambda$ the viscous drag coefficient.}, we propose and experimentally test the relation 
\begin{equation}
D \approx \tau_aT_e
=\frac{E_k}{m(\nu/a^{2}+\gamma_b)}, 
\label{diffusionscaling}
\end{equation}
where $\tau_a$ is the characteristic dissipative time at the scale $1/a$, $\tau_a= \tau(k=1/a)$, and $T_e$ is the usual granular temperature $T_e = \langle v^2\rangle/2 = E_k/m$. Indeed, by analogy with the physics of molecular systems~\cite{Barrat2005,Andreotti2013}, for granular gases the kinetic energy per particle is often expressed in terms of an effective granular temperature $T_e= E_k/m$. The relation Eq.~\ref{diffusionscaling} works especially well in our experiments for $\varepsilon < 10$, when $E_k$ increases with $\varepsilon$ \red{, as shown in Fig. \ref{fig3} (f). The measured value of $D$ is close to $\tau_a\,T_e$ until a turning point at its maximal value corresponding to $\varepsilon =7.30$. However,} for $ \varepsilon >10$, $D$ does not verify as well this scaling law anymore. This result suggests that the Einstein relation (which is a particular case of the fluctuation-dissipation theorem) holds in granular gases with dissipative collisions (for our system $ \varepsilon \lesssim 10$), when the dissipation coefficient is estimated from the velocity correlation functions at the scale $k=1/a$ of few sphere sizes. When the hexagonal crystal phase is approached ($\varepsilon = 62$), we note that $D$ nearly vanishes, which is consistent to the transformation of a fluid-like phase into a solid-like phase.

\section{Deviation from energy equipartition}

\begin{figure*}[htbp]                    
  \begin{center}
  \includegraphics[width=1.7\columnwidth]{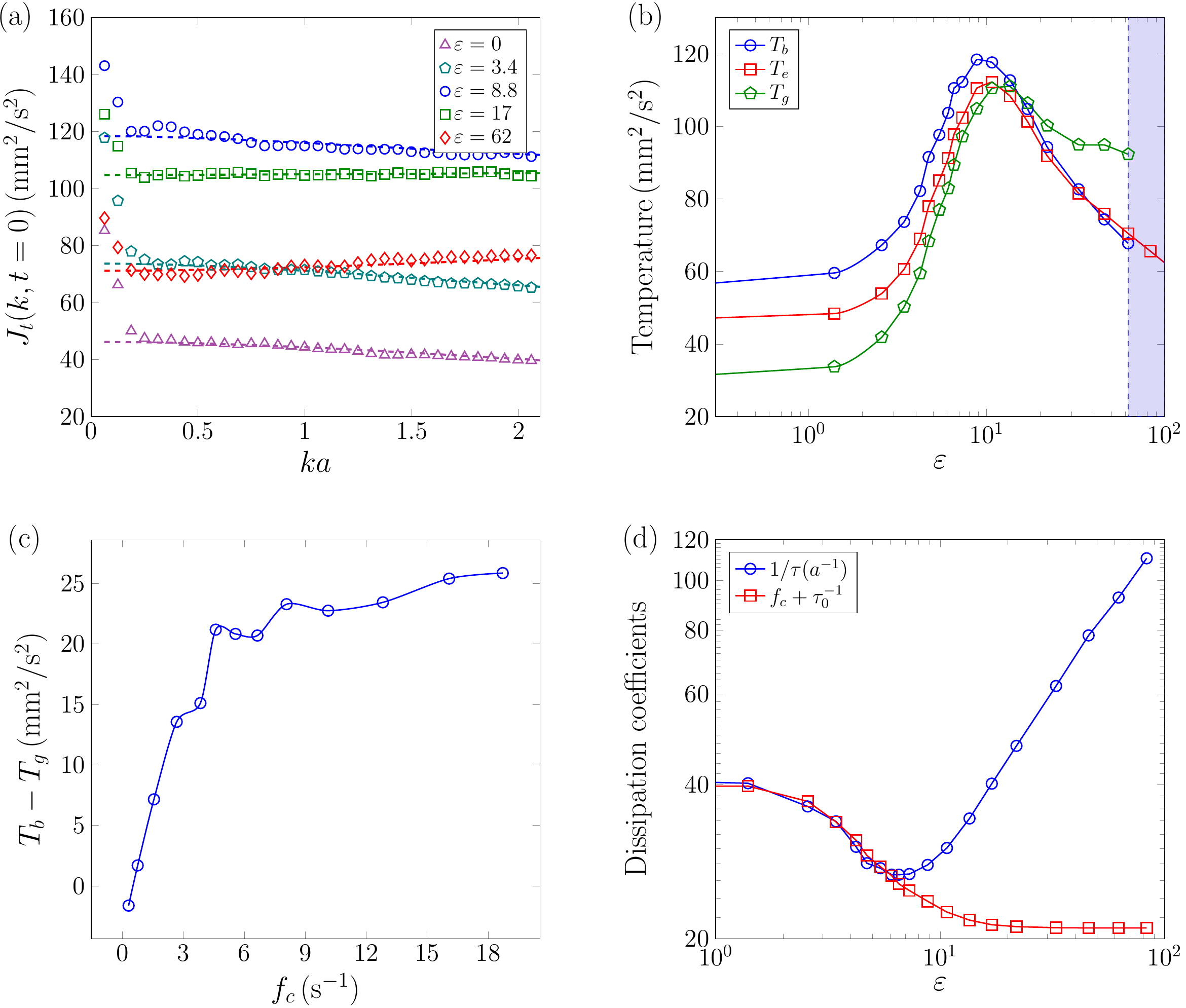}
  \caption{(a) Static transverse velocity structure factor $J_t (k,t=0)$ for selected values of $\varepsilon$. The dashed lines are the fits obtained using Eq.~(\ref{Berhanu_HydroTheo}). (b) Comparison of the usual granular temperature $T_e=E_k/m$, the ``bath temperature" $T_b$, and the ``granular temperature" at the particle scale $T_g$ as a function of $\varepsilon$. The shaded region indicates the transition to the hexagonal phase. (c) Distance to kinetic energy equipartition $T_b-T_g$ as a function of the collision rate $f_c$. (d) Inverse of the dissipative time, $1/\tau_a$ and $f_c$ shifted of an arbitrary constant $\tau_0^{-1} = 21$ s$^{-1}$ as a function of $\varepsilon$.}
    \label{fig4}             
  \end{center}
 \end{figure*}

The static transverse velocity structure factor, $J_t (k,t=0)$, provides the distribution across the spatial scales $k$ of the kinetic energy associated to the transverse modes. Hence, it can be viewed as a kinetic energy power spectrum. A flat spectrum denotes equipartition of energy over the modes. For out-of-equilibrium, dissipative and driven granular gases, the fluctuating hydrodynamics theory defines a ``bath temperature'', $T_b = J_t(k \rightarrow 0,0)$, and a ``granular temperature'' at the particle scale, $T_g = J_t(k \approx 2\pi/a,0)$~\cite{Gradenigotheo2011,Gradenigo2011,Puglisi2012}.  Due to the dissipative collisions acting at the particle scale, $T_b > T_g$. In a system in thermal equilibrium with elastic collisions, these three temperatures $T_e=E_k/m$, $T_b$ and $T_g$ must be equal. In addition, this theory predicts the shape of $J_t (k,0)$ as a function of the characteristic correlation length $\xi=\sqrt{\nu/\gamma_b}$:
\begin{equation}
{J_t (k,0)}={T_g}+\dfrac{{T_b}-{T_g}}{1+\xi^2\,k^2}
\label{Berhanu_HydroTheo}
\end{equation}
Our experimental measurements of $J_t (k,0)$ are plotted for selected values of $\varepsilon$ in Fig.~\ref{fig4} (a). Consistently with the non-monotonic evolution of $E_k$ as a function of $\varepsilon$ (Fig.~\ref{fig1} (c)), the average level of $J_t (k,0)$ increases with $\varepsilon$ until $\varepsilon \approx 0$, then decreases. In the granular gas phase, ($\varepsilon < 62$), $J_t (k,0)$   is well fitted by Eq.~(\ref{Berhanu_HydroTheo}), using the experimentally obtained values of $\xi$, except for the smallest $k$ (Fig.~\ref{fig4}  (a)). The largest peak at $k\approx 0$ may be attributed to a global system vibration rather than to particle dynamics. Note also that since our values of $\xi$ are smaller than those of Puglisi et al.~\cite{Puglisi2012}, $J_t (k,0)$ display less variations than the sigmoidal shapes, that they reported. 
Fitting $J_t (k,0)$ by Eq.~\ref{Berhanu_HydroTheo} provide an estimation of ${T_b}$ and of ${T_g}$, whose difference quantifies the distance to equipartition~\cite{Gradenigotheo2011,Gradenigo2011,Puglisi2012}. In Fig.~\ref{fig4} (b), we compare the temperatures $T_b$, $T_g$ and $T_e$ from the kinetic energy as a function of $\varepsilon$. For moderate $\varepsilon$, we verify that $T_b> T_e > T_g$ and we observe that the distance to equipartition ${T_b}-{T_g}$ decreases with $\varepsilon$ to vanish at $\varepsilon=13.5$ near the maximum of $T_g$. For larger $\varepsilon$, ${T_g}$ is greater than ${T_b}$. The intersection of $T_g$ with $T_b$ corresponds to an inversion of the slope of $J_t (k,0)$ and likely the limit of the validity domain of the fluctuating hydrodynamics theory. For $\varepsilon > 13.5$, the magnetic interactions oppose the large scale fluctuations before inducing crystallization at $\varepsilon \approx 62$. Fluctuating hydrodynamics attributes the difference between ${T_b}$ and ${T_g}$ to dissipative collisions. We verify this statement for $\varepsilon \leq 17$ in Fig.~\ref{fig4} (c). We find indeed that when $f_c$ decreases, $T_b-T_g$ essentially monotonically decreases, to nearly vanish when $f_c=0$. Moreover, the inverse of the dissipative time $1/\tau_a=1/\tau(k=1/a)$, extracted from the time decay of $J_t(k,t)$, is plotted in Fig.~\ref{fig4} (d) as a function of $\varepsilon$.  We find that for $\varepsilon \leq 10$, $1/\tau_a$ is nearly equal to $f_c$ shifted by a positive constant $\tau_0^{-1} = 21\,$s$^{-1}$. Note that $\tau_0^{-1} = 21\,$s$^{-1}$ equals roughly twice the dissipation coefficient $\delta/(1-r^2) \approx 10 $ extracted from the fit of the kinetic energy $E_k$ by $E_k^{th}$ (see Eq.~\ref{Ekmod}). Therefore, the phenomenological dissipation coefficients extracted from the fluctuating hydrodynamics, $\tau_a^{-1}$  reveals that in addition to the collisions between the particles, a supplemental dissipation mechanism, the particles collisions with the bottom and top walls must be also taken into account. The decrease of the particle collision rate, induced by increasing the magnetic field, is precisely reported in $1/\tau_a$ at small $\varepsilon$. For larger values of $\varepsilon$, $1/\tau_a$ significantly increases although $f_c$ nearly vanishes, in correlation with the decrease of $E_k$ and the beginning of crystallization. Our interpretation is that since magnetic repulsion becomes then very strong vertical bead motion is favored, thus reducing the injected power into horizontal motion. 

To summarize, by tuning the collision rate, dissipation can be adjusted although not canceled. The distance to equilibrium can be thus varied, but even at the maximum of the kinetic energy, the system remains out-of-equilibrium and a continuous energy input is needed to maintain a stationary state. However, our study highlights the peculiar role of dissipative collisions between particles in tuning the distance to equipartition. Because they generate dissipation at small scales and are uncorrelated with the forcing (the cell mechanical agitation), these dissipative collisions induce small scale correlations which reduce the kinetic energy spectrum at large $k$. Close to the maximum of the kinetic energy at $\varepsilon \approx 10$, the spectrum is nearly flat,  corresponding to equipartition of the velocity modes. For this quasi-elastic granular gas~\cite{Merminod2014}, spatial structural correlations disappear (Fig.~\ref{fig2} (d)). Additionally, the velocity fluctuations become Gaussian (Fig.~\ref{fig2} (g) (h) and (i)), as for a molecular gas with elastic interactions in thermal equilibrium. For the largest values of $\varepsilon$, the assembly of vibrated spheres becomes structured into a crystalline phase by the magnetic repulsion. Consequently, the kinetic energy at large scale is reduced.

\section{Discussion and Conclusions}

Our experimental study confirms the validity of the fluctuating hydrodynamics theory and extends previous works~\cite{Gradenigo2011,Puglisi2012}. We show a clear relationship between the dissipation of hydrodynamics modes and the rate of dissipative collisions between particles. By tuning this collision rate, we demonstrate that the deviation from the kinetic energy equipartition is a consequence of the small scale dissipation induced by the collisions. When the magnetic energy becomes large compared to the horizontal kinetic energy, \textit{i.e.} $\varepsilon >10$, the limit of the validity domain of the fluctuating hydrodynamics theory is reached. Magnetic repulsive forces modify the system structure, increase its rigidity and induce correlations in the velocity modes. Although a complete description of the system would require to incorporate the magnetic interactions in the fluctuating hydrodynamics, in this work we use a perturbation approach assuming that in first approximation the interactions influence only the collision rate. Our results show the relevance of this hypothesis at least for $\varepsilon \lesssim 10$. Therefore, the fluctuating hydrodynamics theory satisfactorily describes the dynamical properties of such quasi-two-dimensional vibrated granular gas for a large range of collision rates. The influences of the packing fraction and of the agitation strength have been tested by Puglisi et al.~\cite{Puglisi2012}. However, it is a strong hypothesis to equate the energy injection by consecutive particle collisions on the bottom rough wall with a thermal-like noise. Comparisons with other theoretical methods like the mode coupling model~\cite{vanNoijeErnst1999} or with three-dimensional molecular dynamic simulations would be useful to characterize energy injection in quasi-two-dimensional driven granular gas, given the absence of a measurement of particle vertical motions. Finally, we have shown that the out-of-equilibrium specificity of granular gases, unlike molecular gas, is related to the emergence of spatial correlations, here caused by the dissipative collisions. We would encourage to examine similar questions in other out-of-equilibrium systems with heterogeneous energy dissipation or injection such as turbulent flows or assemblies of active particles.

\begin{acknowledgments}
We thank Thierry Hocquet and Martin Devaud for letting us use their computing facilities at MSC, Universit\'e Paris Diderot. We acknowledge Nicol\'as Mujica and Rodrigo Soto from University of Chile for discussions and S\'ebastien Auma\^itre from CEA Saclay, SPEC for comments. Fr\'ed\'eric van Wijland from MSC, Universit\'e Paris Diderot is also gratefully thanked for comments and proofreading of the manuscript. This research was supported by Universit\'e Paris Diderot and Fondecyt Grant No. 3160032 (G.C.).
 \end{acknowledgments}


%

\end{document}